\documentclass[12pt]{iopart}
\usepackage{iopams}
\usepackage{graphicx,color}
\usepackage[colorlinks=true,pdfstartview=FitV,
linkcolor=red,citecolor=green,urlcolor=blue]{hyperref}

\newcommand{\ei}{\end{itemize}}

\newcommand{\Ref}[1]{(\ref{#1})}
\newcommand{\RR}{{\mathbb{R}}}

\newcommand{\NN}{{\mathbb{N}}}
\renewcommand{\aa}{\aalpha}

\newcommand{\aalpha}{{\boldsymbol{\alpha}}}

\newcommand{\av}[1]{\left|{#1}\right|}

\newcommand{\ip}[2]{\left\langle{#1},{#2}\right\rangle}

\renewcommand{\O}{\mathcal{O}}

\newcommand{\R}{\mathsf{R}}
\newcommand{\NR}{\mathsf{NR}}

\newcommand{\x}{{\mathbf{x}}}
\newcommand{\y}{{\mathbf{y}}}

\newcommand{\thet}{{\mathbf{\theta}}}

\newcommand{\Lai}{\Lambda_{\aa,i}}

\newcommand{\lambold}{\boldsymbol{\lambda}}
\newcommand{\fait}{f_{\aa,i}(t)}

\newcommand{\be}{\begin{enumerate}}
\newcommand{\ee}{\end{enumerate}}

\newtheorem{remark}{Remark}
\newtheorem{theorem}{Theorem}
\newtheorem{proposition}{Proposition}

\begin{document}

\title
{Synchronization in networks of general, weakly nonlinear oscillators}

\author{Kre\v simir Josi\' c \dag\ and Slaven Pele\v s \ddag
\footnote[3]{ (slaven.peles@physics.gatech.edu)}
}

\address{\dag\ Department of Mathematics, University of Houston, Houston, TX 77204-3008}

\address{\ddag\ School of Physics, Georgia Institute of Technology,
         Atlanta, GA 30332-0430}

\begin{abstract}
  We present a general approach to the study of synchrony in networks
  of weakly nonlinear systems described by singularly perturbed
  equations of the type $x''+x+\epsilon f(x,x')=0$.  By performing a
  perturbative calculation based on normal form theory we analytically
  obtain an $\O(\epsilon)$ approximation to the Floquet multipliers that
  determine the stability of the synchronous solution.  The technique
  allows us to prove and generalize
  recent results obtained using heuristic approaches, as well as 
  reveal the structure of the approximating equations. We illustrate
  the results in several examples, and discuss extensions to the
  analysis of stability of multisynchronous states in networks with
  complex architectures.  
\end{abstract}

\pacs{05.45.Xt (Synchronization; coupled oscillators), 02.30.Mv (Approximations and expansions)}

\submitto{\JPA}

\maketitle

\section{Introduction}

Networks of coupled oscillators are used to describe a variety of
systems in science and engineering, such as Josephson junction arrays,
generators in power plants, firefly populations, and heart pacemaker
cells. Of particular interest are solutions in which the network, or
subpopulations within the network, oscillate synchronously. The
analysis of the stability of and transition to a synchronous state can
be very complex and has received much attention
\cite{tass,strogatz03,Pik01}. 

Recent applications to 
nanoelectromechanical systems (NEMS) \cite{cross04}, and beam steering 
devices in telecommunications \cite{heath04}, showed that important 
advances can be made by studying these problems perturbatively.  It is 
therefore essential to have appropriate mathematical tools for such an analysis. 
We propose a perturbative method, based on normal form techniques
\cite{nayfeh93,kahn98,deville03}, which is in many respects superior to those commonly 
used to study synchrony in oscillator networks. 

The method is intuitive and helps us distinguish between contributions 
to the dynamics arising from 
the network configuration and the internal structure of individual oscillators. 
Because of this 
it is possible to carry out calculations without having to specify the nonlinearities explicitly.
Also, the approach based on normal forms is rigorous, and the validity of 
approximations is known \emph{a priori}.  
On the other hand, for methods commonly
used in the physics literature, mathematical justification is often
non-trivial, and must be performed \emph{a posteriori} \cite{deville03,murdock03}.

There are a number of other advantages that the normal form approach
brings to the table.  The approximating equations to the
original system are obtained by examining a collection of algebraic
conditions.  This procedure can be formulated in an algorithmic form
and automated, which is of particular importance when
approximations of higher order in the small parameter are needed. 
Computer codes for determining normal forms to any order
already exist for some problems in celestial mechanics. 

Standard methods, such as averaging, usually require center manifold 
reduction to be performed first \cite{carr81}. We will show that the
center manifold reduction is obtained naturally in the normal form of the equations 
of motion. Moreover, the change of coordinates leading to the normal form 
can be used to approximate the center manifold, the invariant fibration
over the center manifolds, and a number of nearly conserved quantities
of the equations to any order.  As a result, the normal form method offers a 
deeper insight into the geometric structure of the approximating equations.

Related approaches can be found in the literature
\cite{kopell86,ashwin}.  In particular, the method of normal forms has
been used in \cite{Hoppensteadt97} to obtain reduced equations for
oscillators close to a bifurcation.  Our approach differs in that we
do not consider only small amplitude oscillations, but general weakly
nonlinear oscillators.  Moreover, the coupling in the present case
results in negative eigenvalues in the linear part of the vector
field.  

\begin{figure}
  \centerline{\includegraphics[width=4cm]{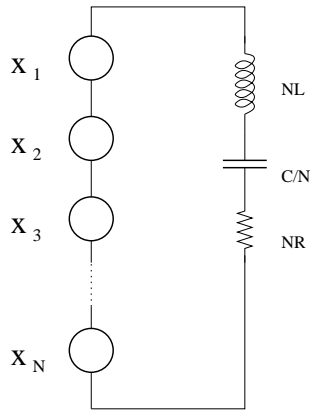}}
  \caption{Globally coupled oscillators. Load parameters are rescaled 
  with respect to the number of oscillators, so the values of 
  load resistance, capacitance and inductance are $NR$, $C/N$ and 
  $NL$, respectively. }
  \label{fig:circuit}
\end{figure}

In this paper we look at systems of globally coupled identical
oscillators illustrated in Figure \ref{fig:circuit}.  This
configuration is commonly used for wave generators in order to
increase the output power (See e.g. \cite{han94}).  
When oscillators are synchronized the
power of emitted waves scales as a square of the number of coupled
units. Therefore, it is important to determine couplings that lead to  
synchronous behavior. Our analysis results in a general expression for
the onset of synchronization in the network, and we recover recent
results for an array of van der Pol oscillators \cite{peles03} as a
special case.  Furthermore, we find, somewhat surprisingly, that the
coupling can induce synchronous oscillations even in a network of
weakly nonlinear systems which are unstable, and do not oscillate when
uncoupled. The method itself can be easily extended to treat more
complex networks and other types of coherent solutions.  In
order to keep our presentation streamlined we do not discuss these
problems here.

This work is motivated by previous studies of synchrony in Josephson
junction arrays \cite{kurt-colet-strogatz96,kurt-swift95,Che95,mukesh02},
where a series of junctions was shunted with
an $RLC$ load (Fig. \ref{fig:circuit}).  
Dhamala and Wiesenfeld introduced a heuristic, perturbative method in
which an approximate stroboscopic map was constructed 
\cite{mukesh02}.  For an appropriately chosen strobing period $T$ the
synchronous solution corresponds to a fixed point of this stroboscopic
map, and the stability of the synchronous state can be determined from
the eigenvalues of its linearization.  A remarkable consequence of
this approach is a unified synchronization law for capacitive and
noncapacitive junctions, two cases which were believed to have
different dynamics.  An extension of the method, and an application to
the study of synchrony in an array of van der Pol oscillators was
given in \cite{peles03}.

The stroboscopic map approach, like most classical singular
perturbation methods, consists of identifying and taming secular terms
in the \emph{naive} approximating solution of the weakly nonlinear
system.  The general structure of the reduced equation obtained
using this approach is difficult to know before the calculations are
carried out.  On the other hand, the normal
form method enables us to carry out calculations without having to
specify the nonlinearity explicitly, or calculate the approximate
strobing time $T$, and hence study a much broader class of problems.

The paper is organized as follows: In Section \ref{S:normal} we
briefly review the theory of normal forms and illustrate it in the
case of a van der Pol oscillator.  These ideas are applied in
Section \ref{S:network} to compute the normal form of the equation
describing a network of weakly nonlinear oscillators.  The stability
of the synchronous solution in this network is analyzed in Section
\ref{S:analysis}.  Several examples illustrating these ideas are given
in Section \ref{S:examples}. Finally, in Section \ref{S:conclusion}
we discuss possible extensions of our work to different networks.

\section{Normal Form Analysis} \label{S:normal}

In this section we give outline the application of the method of normal forms to
the analysis weakly nonlinear systems in a general setting.  The method has been discussed in
\cite{kahn98,nayfeh93}, and a mathematical analysis is given in
\cite{ziane01,deville03}.
Consider a weakly nonlinear ODE of the form
\begin{equation}\label{eq:1}
    \x' = A\x + \epsilon \sum_{\aa,i} c_{\aa,i}\x^\aa e_i = A\x + 
    \epsilon \mathbf f(\x),
\end{equation}
where $\x\in\RR^n$, $A$ is an $n \times n$ constant matrix, and $e_i$ is
the $i$-th unit vector. We use
standard multi-index notation, so that $\aa\in\NN^n$ and $\x^\aa
= x_1^{\alpha_1}\cdots x_n^{\alpha_n}$.  We emphasize that, in contrast
with local normal form theory, $\mathbf f(\x)$ may
contain any monomial in $\x$, including linear terms.

A goal of normal form theory is to remove terms of $\O(\epsilon)$ in
equation \Ref{eq:1} by a near-identity change of variables
\begin{equation} \label{eq:changeofvar}
    \x = \y + \epsilon \mathbf g(\y), \qquad \mathbf g: \RR^n \rightarrow \RR^n,
\end{equation}
where $\mathbf g(\y)$ is a polynomial. In terms of the new variables (\ref{eq:1}) 
becomes
\begin{equation} \label{E:expansion}
     \y' = A\y + \epsilon \left( \sum_{\aa,i}
    c_{\aa,i} \y^\aa e_i - [A,\mathbf g](\y) \right) + \O(\epsilon^2)
\end{equation}
where the Lie bracket $[A,\mathbf g](\y)$ equals $Dg(\y) A\y -A\mathbf g(\y)$.
To remove the nonlinear
terms at $\O(\epsilon)$ in (\ref{E:expansion}), we need to solve the equation
\begin{equation}\label{eq:homologicalsum}
    [A,\mathbf g](\y) = \sum_{\aa,i}c_{\aa,i}\y^\aa
    e_i.
\end{equation}
Since this equation is linear in $\mathbf g$, it is equivalent to the
finite family of equations
\begin{equation}\label{eq:homologicalLB}
    [A,g_{\aa,i}](\y) =  c_{\aa,i}\y^\aa e_i.
\end{equation}
If $A$ is diagonal, the eigenvectors of $[A,\cdot]$ are the
homogeneous polynomials, since
\begin{equation*}
    [A, \y^\aa e_i] = \Lambda_{\aa,i} \y^\aa e_i,
\end{equation*}
where
\begin{equation} \label{E:evalues}
\Lambda_{\aa,i} = \sum_k \alpha_k \lambda_k -\lambda_i =
\ip\aa\lambold - \lambda_i,
\end{equation}
and $\lambda_i$ are eigenvalues of matrix $A$. 
It follows that, if  $ \Lambda_{\aa,i} \neq 0$,
equation \Ref{eq:homologicalLB} has the solution
$$
  g_{\aa,i}(\y) = \frac{c_{\aa,i}}{ \Lambda_{\aa,i}} \y^\aa e_i.
$$
On the other hand, if  $ \Lambda_{\aa,i} = 0$, equation
\Ref{eq:homologicalLB} does not have a solution, and we say that the
monomial $\y^\aa e_i$ is \emph{resonant}. Therefore, only
nonlinear monomials $\y^\aa e_i$ such that  $ \Lambda_{\aa,i} \neq 0$
can be removed at first order in $\epsilon$ by a near identity
coordinate change of the form \Ref{eq:changeofvar}.

In particular, we can split the nonlinearity $\mathbf f(\y)$ in \Ref{eq:1}
into a resonant part $\mathbf f^\R(\y) = \sum_{\Lambda_{\aa, i} = 0}
c_{\aa,i}\y^\aa e_i$ and a nonresonant part $ \mathbf f^\NR(\y)=
\sum_{\Lambda_{\aa, i} \neq 0} c_{\aa,i}\y^\aa e_i$, so that $\mathbf f(\y) =
\mathbf f^\R(\y) + \mathbf f^\NR(\y)$. By setting 
$\mathbf g(\y) = \sum_{\Lambda_{\aa, i} \neq  0} g_{\aa,i}(\y)$, 
the change of coordinates \Ref{eq:changeofvar} leads to the equation
\begin{equation} \label{E:reduced}
 \y' = A\y + \epsilon \mathbf f^\R(\y) + \O(\epsilon^2).
\end{equation}

We emphasize that to obtain the normal form of equation \Ref{eq:1} to 
$\O(\epsilon)$, we simply identify and remove all resonant terms, that is all monomials
comprising $\mathbf f^\NR(\x)$.  The preceding argument shows that this can be
done at the expense of introducing terms of $\O(\epsilon^2)$ into the equation.  If the $\O(\epsilon^2)$ 
terms are neglected in \Ref{E:reduced}, a \emph{truncated normal form} is obtained.  
To continue this process and obtain normal forms to higher order in $\epsilon$, it is 
necessary to compute the  $\O(\epsilon^2)$ terms that are introduced at this step explicitly.  
This is equivalent to the observation that the computation of a \emph{local} normal form
near a singular point to second order may affect the cubic terms, and the computation
needs to be carried out order by order (see \cite{murdock03}, for instance).

The following theorem shows that the truncated normal form provides a
good approximation to the original equations

\begin{theorem}\cite{deville03} \label{thm:stayclose} Consider the ordinary differential
equation
\begin{equation*}
    \x' = A\x + \epsilon \sum_{\aa,i}\fait \x^\aa e_i, \qquad \x(0) = \x_0
\end{equation*}
where $A$ is a diagonal, and has eigenvalues with non-positive real
part.  Construct the truncated  normal form
\begin{equation*}
    \y' = A\y + \epsilon \mathbf f^\R(\y),
\end{equation*}
and let $\x(0)=\y(0)$. Then there is a $T=T(\x(0))>0$ such
that the solutions of the two equations satisfy
$\av{\x(t)-\y(t)}=\O(\epsilon)$ for all $t\in[0,T/\epsilon]$ and $\epsilon$
sufficiently small.
\end{theorem}

\begin{remark}  When the truncated equations contain hyperbolic
  invariant structures, the conclusions of Theorem \ref{thm:stayclose}
  often hold for all time along their stable directions
  \cite{hale80}.  Normal form theory can be extended to study
  nondiagonalizable matrices \cite{murdock03}, nonhomogeneous
  equations with nonlinearities that are not finite sums of monomials,
  and higher order approximations in $\epsilon$
  \cite{deville03,mitropolsky95}.  The main ideas presented here are
  similar in these cases. We present the simplest case in order to
  keep technicalities at a minimum.
\end{remark}

Note that, by construction $[A, \mathbf f^\R] = 0$.  It follows that the
truncated normal form is equivariant under the flow of the
unperturbed equation $\x' = A \x$.  Moreover, the resonant monomials
and hence the structure of the truncated normal form, are completely
determined by the eigenvalues of $A$.  This allows us to prove the
following Proposition which will be useful in the following sections.  

\begin{proposition} \label{P:remove1} Suppose that matrix $A$ in 
(\ref{eq:1}) is diagonalizable, has $m$ purely imaginary eigenvalues $\lambda_i$, and 
that other eigenvalues $\nu_i$ have negative real part. The system 
(\ref{eq:1}) then can be written as
\begin{eqnarray}
\x_1' &=& A_1\x_1 + \epsilon \mathbf h_1(\x_1,\x_2)  \nonumber\\
\x_2' &=& A_2\x_2 + \epsilon \mathbf h_2(\x_1,\x_2) \nonumber
\end{eqnarray}
where $A_1 = diag(\lambda_1,\ldots, \lambda_m)$, 
$A_2= (\nu_{m+1}, \ldots \nu_n)$.  If the nonlinear terms 
$\mathbf h_1$ and $\mathbf h_2$ are polynomials, the truncated normal form of this
system has the general form 
\begin{eqnarray}
\y_1' = A_1\y_1 + \epsilon \mathbf h_1^{\R}(\y_1) \label{E:xeqn} \\
\y_2' = A_2\y_2 + \epsilon \mathbf h_2^{\R}(\y_1,\y_2) \label{E:yeqn}
\end{eqnarray}
where $\mathbf h_2^{\R}(\y_1,0) = 0$.
\end{proposition}

{\bf Proof:} 
A term $\x_1^\aa \x_2^{\boldsymbol{\beta}} e_i$ with $i\leq m$ 
is resonant if
$\Lambda_{\aa,i} = \sum_{j=1}^m \alpha_j \lambda_j + \sum_{j=m+1}^n
\beta_j \nu_j - \lambda_i = 0$.  Since the eigenvalues $\nu_i$ have negative
real part and enter the sum with the same sign, this condition can be
satisfied only if all $\beta_j = 0$.

Similarly, a term $\x_1^\aa e_i$ for $i > m$ is resonant only if 
$\Lambda_{\aa,i} = \sum_{j=1}^M \alpha_j \lambda_j - \nu_i = 0$.  This
equation cannot hold. Hence all resonant monomials in \Ref{E:yeqn}
contain a nonzero power of $y_{2,j}$ for some $j$, and evaluate to 0 when
$\y_2 = 0$.   
$\square$

Although Proposition \ref{P:remove1} is simple to prove, it says much
about the structure of the truncated normal form.  The fact that
$\mathbf h_2^{\R}(\y_1,0) = 0$ means that the hyperplane $\y_2 = 0$ is invariant
under the flow of (\ref{E:xeqn}-\ref{E:yeqn}).  In fact, the
hyperplane $\y_2 = 0$ is the center manifold of this system, and hence
\Ref{E:xeqn} trivially provides the reduction of the truncated normal
form equation to the center manifold.  Therefore, it is unnecessary to  
compute the center manifold explicitly to obtain the reduced
equations.

Furthermore, since $\y_2$ does not occur on the right hand side of
\Ref{E:xeqn}, the fibration given by $\y_1 = const.$ is also invariant
under the flow.  To obtain an $\O(\epsilon)$ approximation of the
center manifold, and the invariant fibration over the center manifold
in the original coordinates, it is sufficient to invert the near
identity transformation \Ref{eq:changeofvar} used in obtaining the
normal form equation.  In fact, there are typically other easily
identifiable quantities that are conserved by the flow of the
truncated normal form, and provide adiabatic invariants for the
original equations \cite[Chapter 5]{murdock03}.

\subsection{Example: Van der Pol Oscillator}\label{subsec:vdp}

As a simple, illustrative example, and to introduce results that will
be used in section \ref{S:examples}, we consider the van der Pol
equation
\begin{equation}
  \label{eq:vdp}
  x'' - \epsilon x' (1-x^2) + x = 0.
\end{equation}
We can rewrite system (\ref{eq:vdp}) in the variables
\begin{equation}
  \label{eq:zz}
  z = x + i x', ~~~\bar{z} = x - i x'
\end{equation}
to obtain:
\begin{eqnarray}
  \label{eq:vdpz1}
  z' & = & -i z +\frac{\epsilon}{8}
  (4z - 4\bar{z} -z^3 -z^2 \bar{z} + z\bar{z}^2 + \bar{z}^3),
   \\
  \bar{z}' & = & i \bar{z} - \frac{\epsilon}{8}
  (4z - 4\bar{z} -z^3 -z^2 \bar{z} + z\bar{z}^2 + \bar{z}^3). \label{eq:vdpz2}
\end{eqnarray}
This system is in the form
that can be analyzed using the ideas discussed in Section
\ref{S:normal}.  The eigenvalues of the linear part are $\lambda_1=-i$
and $\lambda_2=i$. The resonant terms can be computed using the
condition $\Lai = 0$ where $\Lai$ is defined in equation
\Ref{E:evalues}.

\begin{table}[htbp]
  \caption{\label{tab:resonance} Resonance condition.}
  \begin{indented}
  \item[]
  \begin{tabular}{ccccccc}
    \br
    term & $z$ & $\bar z$ & $z^3$ & $z^2$ $\bar z$ &
    $z \bar{z}^2$ & $\bar{z}^3$ \\
    \mr
    $\boldsymbol{\alpha}$& (1,0)& (0,1)& (3,0)& (2,1)& (1,2)& (0,3) \\
    $\Lambda_{\boldsymbol{\alpha},1}$& 0 & $2i$ & $-2i$ & 0 & $2i$ & $4i$\\
    $\Lambda_{\boldsymbol{\alpha},2}$ & $-2i$ & 0 & $-4i$ & $-2i$ & 0 & $2i$\\
    \br
  \end{tabular}
  \end{indented}
\end{table}

As noted in the discussion following the derivation of equation \Ref{E:reduced}, 
the truncated normal can be obtained simply by removing the resonant monomials 
in (\ref{eq:vdpz1}-\ref{eq:vdpz2}).
From Table \ref{tab:resonance} we find that the resonant terms in
\Ref{eq:vdpz1} are $z$ and $z^2\bar z$, and the resonant terms in
\Ref{eq:vdpz2} are their complex conjugates $\bar z$ and $z\bar{z}^2$,
as expected.  Therefore, there exists a near identity change of
coordinate in which    (\ref{eq:vdpz1}-\ref{eq:vdpz2}) have the form
\begin{eqnarray}\label{eq:vdpnormal}
  z' & = & -i z +\frac{\epsilon}{2}z  -\frac{\epsilon}{8}z^2
  \bar{z}+\O(\epsilon^2), 
\end{eqnarray}
and it's complex conjugate.  

As noted in Section \ref{S:normal}, equation \Ref{eq:vdpnormal} is
equivariant under the flow of the unperturbed equation, which is a
pure rotation of the real plane.  It follows that the right hand side
of any weak perturbation of the equation $z' = -i z, \bar{z}' =
iz$ cannot depend on the angular variable when expressed in polar
coordinates.  Indeed, \Ref{eq:vdpnormal} takes
the form
\begin{equation*}
R' =  \frac{\epsilon}{2} R \left( 1 - \frac14 R^2 \right), 
\qquad \qquad \theta' = 1.
\end{equation*}
in polar coordinates.  The same procedure can be used to obtain
higher order normal forms, see \cite{deville03}.

\begin{remark} \label{rem:newcoordinates} 
  Strictly speaking, new coordinates are introduced in obtaining
  equation \Ref{eq:vdpnormal}.  To keep notation at a minimum, we name
  these new variables $z$ and $\bar{z}$ as well.  A similar convention
  is used in the rest of the paper.
\end{remark}

\section{Globally coupled networks} \label{S:network}

In this section we use the normal form method to study a network of
identical, weakly nonlinear oscillators, described by the equation
$x_i''+x_i+\epsilon h(x_i,x'_i)=0$ when uncoupled.  Here $\epsilon$ is
a small parameter, and the nonlinear term $h(x_i,x'_i)$ is
assumed to be a polynomial. The elements in the network are globally
coupled by a linear load (Fig.  \ref{fig:circuit}). The coupling is
weak and of the same order as the nonlinearity. While this is not the
most general example of a globally coupled network, these assumptions
have been chosen to simplify the presentation, and can be relaxed.
Equations of motion for this system can be written as
\begin{eqnarray}
  \label{eq:osc}
  x_k'' + \epsilon h(x_k,x_k') + x_k  =  q' \\
  \label{eq:load}
  \mu_1 q'' + \mu_2 q' + q  =
  \epsilon \frac{\kappa}{N}\sum_{j=1}^{N}x_j
\end{eqnarray}
Here we follow the notation introduced in
\cite{Che95,kurt-colet-strogatz96,mukesh02,peles03}, where 
$\mu_1$ and $\mu_2$ are control parameters, and can be understood as 
the inductance and the resistance of the coupling load, respectively.  
The goal of our calculation is to find load parameters that will 
ensure synchrony in the network. 
            
In order to bring the equations of motion to normal form, we first
have to diagonalize their linear parts. To make the procedure more
intuitive we introduce complex variables $z_k = x_k + i x'_k$ and
$\bar{z}_k = x_k - i x'_k$, and denote $q'=p$. The system
(\ref{eq:osc}-\ref{eq:load}) then becomes
\begin{eqnarray}
  \label{eq:osczz}
  z_k'  =  - & i  z_k + i p -i \epsilon f(z_k,\bar{z}_k) \\
  \bar{z}_k'  =  & i \bar{z}_k - i p +i \epsilon f(z_k,\bar{z}_k)
\end{eqnarray}
where $f(z, \bar{z}) = h[(z+\bar{z})/2,(z-\bar{z})/2]$, and
\begin{eqnarray}
  q'  =  p \\
  \label{eq:loadzz}
  p'  =  -\frac{\mu_2}{\mu_1} p -\frac{1}{\mu_1} q
  + \epsilon \frac{\kappa}{2 \mu_1 N}\sum_{j=1}^{N}(z_k + \bar{z}_k). 
\end{eqnarray}
This system can be written in matrix form as
\begin{equation}
  \label{eq:matofmotion}
  {\bf z}' = A {\bf z} + i \epsilon {\bf f}({\bf z})
\end{equation}
where ${\bf z}=(z_1,\bar{z}_1,z_2\ldots,\bar{z}_N,q,p)$,
\begin{equation}
  A= \left(
    \begin{array}{cccccccc}
      - i & 0 & 0 & 0 &  & 0 & 0 & i\\
      0 & i & 0 & 0 &  & 0 & 0 & - i\\
      0 & 0 & - i & 0 &  & 0 & 0 & i\\
      0 & 0 & 0 & i &  & 0 & 0 & - i\\
      &   &   &   & \ddots &  &  & \vdots\\
      0 & 0 & 0 & 0 &  & i & 0 & - i\\
      0 & 0 & 0 & 0 &  & 0 & 0 & 1\\
      0 & 0 & 0 & 0 &  & 0 & - \frac{1}{\mu_1} & -\frac{ \mu_2}{\mu_1}
    \end{array} \right) \nonumber
\end{equation}
and
\begin{equation}
\fl
  {\bf f}({\bf z})= \left(
      -f(z_1,\bar{z}_1),
      f(z_1,\bar{z}_1),
      -f(z_2,\bar{z}_2),
      \ldots,
      f(z_N,\bar{z}_N),
      0,
      - \frac{i \kappa}{2 \mu_1 N}\sum_j(z_j+\bar{z}_j) \right)^T. \nonumber
\end{equation}
Here $^T$ denotes the transpose, and ${\bf f}({\bf z})$ is a column vector.

Finally, we diagonalize the matrix $A$, 
by introducing another coordinate change ${\bf w} = B^{-1}{\bf z}$, where 
$B^{-1}AB = D_A$ is diagonal.  In the new coordinates the
equations of motion have the form
\begin{equation}
  \label{eq:normform}
  {\bf w}'  =  B^{-1}A{\bf z} + i \epsilon B^{-1}{\bf f}({\bf z})
  =  D_A {\bf w} + i \epsilon B^{-1}{\bf f}(B{\bf w}),
\end{equation} 
The matrix $B$ can be computed using elementary linear algebra.
Its actual form is not of interest, and we therefore suppress it. 
In component form (\ref{eq:normform}) becomes
\begin{eqnarray}
  \label{eq:normform-components1}
  w_k' & = & -i w_k
  -i\epsilon f(w_k+b_1 u + b_2 v,\bar{w}_k+\bar b_1 u + \bar b_2 v)\\
  & & + \epsilon \frac{\kappa}{2 Z N}e^{i \delta}
  \sum_j[w_j+\bar{w}_j+v(b_2+\bar b_2)+u(b_1+\bar b_1)]\nonumber \\
  \label{eq:normform-components2}
  \bar{w}_k' & = & i\bar{w}_k
  +i\epsilon f(w_k+b_1 u + b_2 v+\bar{w}_k+\bar b_1 u + \bar b_2 v)\\
  & & + \epsilon \frac{\kappa}{2 Z N}e^{-i \delta} 
  \sum_j[w_j+\bar{w}_j+v(b_2+\bar b_2)+u(b_1+\bar b_1)]\nonumber \\
  \label{eq:normform-components3}
  u' & = & \nu_{2N+1} \, u \\ & & -i(1 - C) 
  \frac{\kappa}{4\mu_1 N}\sum_j[w_j+\bar{w}_j+v(b_2+\bar b_2)+u(b_1+\bar b_1)]\nonumber \\
  \label{eq:normal-last}
  v' & = & \nu_{2N+2} \, v  \\  & & -i(1 + C)
  \frac{\kappa}{4\mu_1 N}\sum_j[w_j+\bar{w}_j+v(b_2+\bar b_2)+u(b_1+\bar b_1)] \nonumber
\end{eqnarray}
where $b_{1,2} = 2\mu_1/(2\mu_1 +i \mu_2 \mp i\sqrt{\mu_2^2 - 4 \mu_1})$,
$C=\mu_2/\sqrt{\mu_2^2-4\mu_1}$, $Z=\sqrt{(1-\mu_1)^2+\mu_2^2}$ is
the impedance, and $\delta = \arcsin (\mu_2/Z)$ is the phase shift on
the load.
Variables $u$ and $v$ are obtained from $q$ and $p$.
Note that matrix $A$ has $2N$ purely imaginary eigenvalues
$\lambda_k = -i, \lambda_{k+1} = i, k = 1, 3, \ldots, 2N-1$,
corresponding to the first $2N$ entries on the diagonal of $B^{-1}AB$. 
The last two eigenvalues
\begin{equation}
  \label{eq:nu12}
  \nu_{2N+1,2} = \frac{-\mu_2 \pm \sqrt{\mu_2^2 -4\mu_1}}{2 \mu_1}
\end{equation}
have negative real part. 

Since the linear part of (\ref{eq:normform-components1}--\ref{eq:normal-last})
is diagonal, it is now straightforward 
to identify nonresonant terms, as discussed in Section \ref{S:normal}.
First, from Proposition \ref{P:remove1}, we find that all terms
containing powers of $u$ or $v$ can be removed at $\O(\epsilon)$ in
the equations for $w_k$ and $\bar{w}_k$ using a near identity change
of coordinates, so (\ref{eq:normform-components1}-\ref{eq:normform-components2}) 
reduce to the equations on the center manifold 
\begin{eqnarray}
  \label{eq:ww}
  w_k'  =  -& i w_k -i\epsilon f(w_k,\bar{w}_k)
  + \epsilon \frac{\kappa}{2 Z N}e^{i \delta}
  \sum_j(w_j+\bar{w}_j)
  + \O(\epsilon^2),\\
  \bar{w}_k'  = & i\bar{w}_k +i\epsilon f(w_k,\bar{w}_k)
  + \epsilon \frac{\kappa}{2 Z N}e^{-i \delta}
  \sum_j(w_j+\bar{w}_j)
  + \O(\epsilon^2).\label{eq:ww2}
\end{eqnarray}
While the load variables do not enter the final
approximating equations, the form of the load equation was important
in obtaining the diagonalization, and will therefore be reflected in
the final approximation.  

We can use the normal form to compute the transients, as well as the
asymptotic state of a solution.  However, since we are interested in
the stability of the synchronous state, we only consider the reduced
equations on the center manifold $u=0$, $v=0$, and therefore drop
equations (\ref{eq:normform-components3}-\ref{eq:normal-last}) from
further consideration.

The normal form of
(\ref{eq:normform-components1}-\ref{eq:normform-components2})
is obtained by computing the remaining resonant terms. 
The ones due to the nonlinearity $f$, are determined as follows:
Let $f^\R_w$ be the resonant part of
$f$ in the equation for the individual oscillators, 
$w' = -i w - i \epsilon f(w, \bar{w})$,
and $f^\R_{\bar{w}}$ the resonant part in
$\bar{w}'= i \bar{w} + i \epsilon \bar f(w,\bar{w})$.
A simple computation shows that
$f^\R_w = w_k\phi^{\R}(w_k \bar w_k)$ and
$f^\R_{\bar{w}_k} = \bar w_k \bar \phi^{\R}(w_k \bar w_k)$,
where $\phi^{\R}$ is a polynomial in $w\bar{w}$.
Generally, the coefficients of $\phi^{\R}$ are complex.

It remains to determine the resonant terms that are due to the coupling
term $\kappa e^{i\delta}/(2 Z N)\sum_j(w_j+\bar{w}_j)$. 
Obviously, monomials $w_j$ are resonant in 
\Ref{eq:ww}, while monomials $\bar{w}_j$ are resonant in equations \Ref{eq:ww2}.
Keeping only the resonant terms in the (\ref{eq:ww}--\ref{eq:ww2}),
we obtain the normal form to $\O(\epsilon)$:
\begin{eqnarray}
  \label{eq:wwnormal1}
  w_k'  =  - & i w_k +\epsilon \left( w_k\phi^{\R}(w_k \bar w_k)
  + \frac{\kappa}{2 Z N}e^{i\delta}\sum_jw_j \right)
  + \O(\epsilon^2)\\
  \label{eq:wwnormal2}
  \bar{w}_k'  = & i\bar{w}_k + \epsilon \left(\bar w_k \bar \phi^{\R}(w_k \bar w_k)
  +  \frac{\kappa}{2 Z N}e^{-i\delta}\sum_j\bar{w}_j \right)
  + \O(\epsilon^2).
\end{eqnarray}

Since the normal form equations are equivariant under the flow of the
unperturbed system, it is again natural to rewrite them in polar
coordinates, $w_k=r_k e^{-i\theta_k}$.  We obtain
\begin{eqnarray}
  \label{eq:polarnormal}
  r_k' &=& \epsilon r_k R(r_k)
  +\epsilon \frac{\kappa}{2ZN}\sum_j r_j \cos(\theta_k-\theta_j+\delta)
  +\O(\epsilon^2)\\
  \theta_k' &=& 1 -\epsilon \Theta(r_k)
  -\epsilon \frac{\kappa}{2ZN}
  \sum_j \frac{r_j}{r_k} \sin(\theta_k-\theta_j+\delta) +\O(\epsilon^2),
\label{eq:polarnormalth}
\end{eqnarray}
where $R=(\phi^{\R}+\bar \phi^{\R})/2$ is the real, and
$\Theta=(\phi^{\R}-\bar \phi^{\R})/2i$ the imaginary part of
$\phi^{\R}$.  
Note that the fact that
the right hand side of \Ref{eq:polarnormal} depends only on the phase
differences is a consequence of the equivariance of the truncated
normal form under the transformation $\thet \rightarrow \thet + C$, where 
$\thet = (\theta_1, \theta_2, \ldots, \theta_N)$ and $C = (c, c, \ldots,
c)$.

\begin{remark}
  We could also use a center manifold reduction to remove the
  variables $u$ and $v$ from equations
  (\ref{eq:normform-components1}-\ref{eq:normform-components2}) \cite{carr81}.
  This would add another step to the calculation.
If the transient dynamics of initial states
off the center manifold is of interest, in addition it is necessary to 
compute the stable fibration over the center manifold.  The normal form
method considerably simplifies these computations. As noted in Section 
\ref{S:normal}, the truncated normal form is a skew product, and provides both 
the reduction of the equations to the center manifold, and an approximation for the flow 
in the transversal direction \cite{cox95}.
\end{remark}

\section{Existence and stability of the synchronous state} \label{S:analysis}

The truncated normal form obtained from
(\ref{eq:polarnormal}-\ref{eq:polarnormalth}) by neglecting
$\O(\epsilon^2)$ terms is much easier to analyze than the original
equations (\ref{eq:osc}-\ref{eq:load}).  It is straightforward to find
the inphase solution for the truncated system and its stability
analytically.  Furthermore, if the truncated system has a stable
synchronous solution, so does the full system
(\ref{eq:osc}-\ref{eq:load}).

Due to all-to-all coupling, the truncated normal form given by
(\ref{eq:polarnormal}-\ref{eq:polarnormalth}) is equivariant under all
permutations of the oscillators, as well as the the action of the
group $T^1$ given by $\theta_i \rightarrow \theta_i + C$ for all $i$.
As a consequence, a number of phase locked states are forced to exist
\cite{ashwin}.  In this section we consider the stability of the synchronous
state $r_1 = r_2 = \ldots r_N = r$, and $\theta_1 = \theta_2 = \ldots =
\theta_N$.  The stability of other phase locked states can be analyzed
similarly.

Assuming that $r_k=r$, and  $\theta_k - \theta_j = 0$ for all $k$ and
$j$, then equation (\ref{eq:polarnormal}) implies that $r$ must satisfy
\begin{equation}
  \label{eq:fixed}
  rR(r) + r\frac{\kappa}{2 Z} \cos \delta = 0.
\end{equation}

\begin{remark} The amplitudes of the uncoupled
  oscillators are given by solutions of $R(r) = 0$, while in a
  synchronously oscillating network the amplitudes of the oscillators
  are given as solutions of \Ref{eq:fixed} in terms of the coupling
  strength $\kappa$, and properties of the coupling load.  It
  is possible that $R(r) = 0$ has only $r = 0$ as a solution, while
  \Ref{eq:fixed} has nonzero solutions for certain values of $\kappa,
  Z, \delta$.  In such examples, the uncoupled systems do not
  oscillate, while the network can exhibit synchronous oscillations
  (see Section \ref{S:sinks}). \label{R:sinks}
\end{remark}

Due to the $T^1$ equivariance of the system, the Jacobian is constant
along the synchronous solution. Therefore the Floquet exponents equal
the eigenvalues of the Jacobian.  At the synchronous solution
determined by $\theta_k - \theta_j = 0$ and \Ref{eq:fixed} they equal
\begin{eqnarray}
  \lambda_1 & = &  0 \\
  \label{eq:lambda1}
  \lambda_2 & = & \epsilon r R'(r) \\
  \label{eq:lambda23}
  \lambda_{n,n+1} & = & \epsilon r R'(r)- \epsilon \frac{\kappa}{Z}\cos \delta \\
  & & \pm \epsilon \sqrt{(r R'(r))^2-\left(\frac{\kappa}{Z}\sin \delta \right)^2
      +2 r \Theta'(r) \frac{\kappa}{Z}\sin \delta}\nonumber
\end{eqnarray}
where $n=3,5,7,\ldots, 2N-1$.  If $r$ satisfies \Ref{eq:fixed} and the
$\lambda_{i>1}$ have non-zero real part then the system obtained by
truncating (\ref{eq:polarnormal}-\ref{eq:polarnormalth}) has a
\emph{weakly hyperbolic} limit cycle with $r_i = r$ and $\theta_i =
\theta_j$ for all $i$ and $j$.  If the limit cycle is stable the
  truncated system is synchronized.  Eigenvalues
  (\ref{eq:lambda1}-\ref{eq:lambda23}) are expressed in terms of the
  coupling load parameters, and implicitly define the region
  corresponding to stable synchronous behavior in parameter
  space.  
  
It remains to show that the stability of the synchronous 
solution in the truncated system implies the existence and 
stability of nearby synchronous solution in the original system 
(\ref{eq:osc}-\ref{eq:load}).  Let $\Delta_{j-1,j} = \theta_{j} -
\theta_{j-1}$ for $j = 2, \ldots, N$, and let $\chi = (\Delta_{1,2},
\ldots, \Delta_{N-1,N}, r_1, \ldots, r_N)$.  The only $\O(1)$ terms
in equations (\ref{eq:polarnormal}-\ref{eq:polarnormalth}), are the unit terms 
in (\ref{eq:polarnormalth}).  By definition of $\Delta_{j-1,j}$, these terms
do not occur in the differential equation for $\chi$.  It follows that that 
in the new coordinates $(\chi, \theta_1)$, equations (\ref{eq:polarnormal}-\ref{eq:polarnormalth}) have the 
form
\begin{eqnarray} \label{eq:singular}
\chi' & = & \epsilon F_1(\chi) + \epsilon^2 F_2(\chi, \theta_1) \nonumber \\
\theta'_1 & = & 1 + \epsilon G_1(\chi) + \epsilon^2 G_2(\chi,\theta_1).
\end{eqnarray}
The explicit form of $F_1, F_2, G_1,$ and $G_2$ is not of importance. 

To study the stability of the synchronous state $\chi_0 = ({\mathbf 0}, {\mathbf
  r})$, we note that the eigenvalues of $\epsilon D_\chi F_1(\chi_0)$ are given
 by (\ref{eq:lambda1}-\ref{eq:lambda23}) when
${\mathbf r} = (r, r, \ldots, r)$ and $r$ satisfies
\Ref{eq:fixed}.  The following proposition shows that if the
eigenvalues $\lambda_i$ have non-zero real part then they completely
determine the stability of the synchronous state for the full system 
(\ref{eq:osc}-\ref{eq:load}).

\begin{proposition}\label{prop:floquet}
  If $F_1(\chi_0) = 0$, and $D_{\chi}F_1(\chi_0)$ has eigenvalues with non-zero
  real part in \Ref{eq:singular}, then there exists an $\epsilon_0$,
  such that system \Ref{eq:singular} has a limit cycle $\O(\epsilon)$ close to the
  limit cycle of the unperturbed system, obtained from
  \Ref{eq:singular} by setting $F_2 = G_2 = 0$, for all $\epsilon <
  \epsilon_0$.  The Floquet exponents of the perturbed limit cycle
  agree with the eigenvalues of $D_{\chi}F(\chi_0)$ to $\O(\epsilon)$.
 \end{proposition}

 {\bf Proof:} This is a consequence of standard results about near 
identity changes of coordinates for systems with a single frequency.
See \cite[Chapter 1.22]{haller99} and references therein. $\square$

\begin{proposition} 
  Assume that a nonzero solution $r$ of equation \Ref{eq:fixed}
  exists.  If the eigenvalues given in
  (\ref{eq:lambda1}-\ref{eq:lambda23}) have negative real part, then
  \Ref{eq:osc} has a stable, synchronous periodic solution.
  If one of the eigenvalues has positive real part, this periodic
  solution is unstable.
\end{proposition}

{\bf Proof:} As a consequence of Proposition \ref{prop:floquet},
depending on the real part of the eigenvalues in
(\ref{eq:lambda1}-\ref{eq:lambda23}) there is either a stable or
unstable synchronous solution of system
(\ref{eq:normform-components1}-\ref{eq:normal-last}).  Since the
change coordinates $B \mathbf{w = z}$ affects all of the pairs of
coordinates $(w_k, \bar{w}_k)$ in the same way, a synchronous solution
in the $w$ coordinates corresponds to a synchronous solution in the
original $z$ coordinates.  The stability properties of this solution
are preserved under a linear change of coordinates. $\square$

\begin{remark} 
  Theorem \ref{thm:stayclose} only states that the truncated normal
  form provides an $\O(\epsilon)$ approximation 
  on timescale of $\O(1/\epsilon)$, and typically this
  approximation does not hold on longer timescales.
  Nevertheless, Proposition \ref{prop:floquet} implies that the
  limit cycle of the truncated normal form approximates the limit
  cycle of the original system to $\O(\epsilon)$, since the
  approximation of the amplitude $\chi$ is valid for all time.
\end{remark}

\begin{remark}
  In \cite{peles03}, a stroboscopically discretized system was used in
  a heuristic argument to obtain similar result in the case of van der
  Pol oscillators.  There it was necessary to determine the angular
  frequency of the inphase solution to $\O(\epsilon)$ in order to find
  appropriate strobing period.  The angular frequency of the periodic
  solution can also be estimated from
  (\ref{eq:polarnormal}-\ref{eq:polarnormalth}) up to second order as
\begin{equation}\label{eq:omega}
\omega =  1 -\epsilon \left( \Theta(r) + \frac{\kappa}{2Z} \sin\delta \right).
\end{equation}
\end{remark}

\section{Examples} \label{S:examples}

\begin{figure}
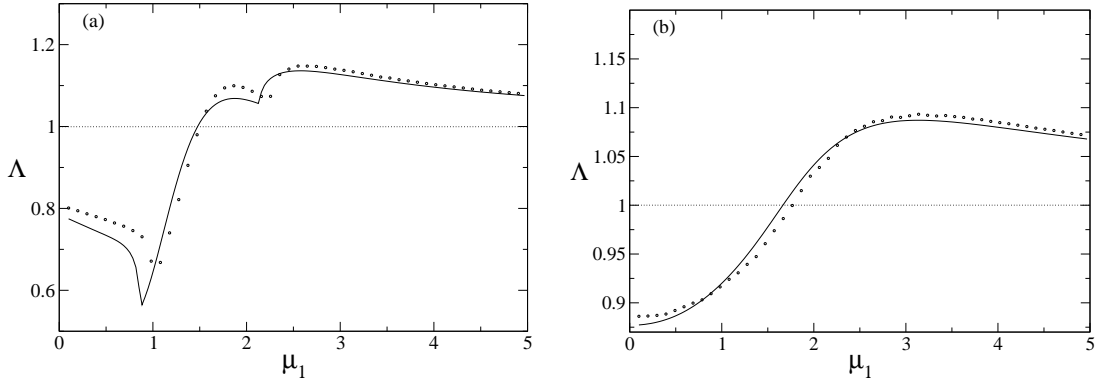

  \centerline{
    \includegraphics[width=7cm]{vdp08}
    \hspace{0.2cm}
    \includegraphics[width=7cm]{vdp15}
  }
  \caption{Floquet multipliers for a van der Pol oscillator
    array. Solid line represents analytical result, while dots represent
    results obtained from numerical calculations. The coupling parameters are
    $\kappa=1$, $\epsilon=0.1$, and
  (a) $\mu_2=0.8$, (b) $\mu_2=1.5$.}
  \label{fig:vdp}
\end{figure}

The synchronization condition for the array of globally coupled oscillators 
is obtained from (\ref{eq:lambda1}-\ref{eq:lambda23}) by setting 
$\lambda_{i}<1$ for all $i>1$, and can be expressed in terms of control 
parameters $\mu_1$ and $\mu_2$. In general, our method allows 
us to do a single calculation for a certain network configuration, and then 
obtain results for a variety of different oscillators in a straightforward 
fashion. To find the synchronization condition for a specific oscillator type 
it suffices to find 
the function $\phi^{\R}$ which characterizes the resonant terms in the
equation of motion of the uncoupled oscillator. 
We illustrate this in several examples and compare 
our approximating solutions with
numerical results. Agreement between analytical and numerical calculation 
is good even when using only first order
corrections in $\epsilon$. We retrieve result from \cite{peles03} as a special 
case of our general result.

\subsection{Van der Pol oscillator arrays}

Consider a network of globally coupled van der Pol oscillators
described in Sec. \ref{subsec:vdp}.
From (\ref{eq:vdpnormal}) we find that the resonant terms are
described by $\phi^{\R} = 1/2 - w \bar w/8$, and
hence $R(r)=1/2-r^2/8$, $\Theta(r)=0$. From (\ref{eq:fixed})
we find that the inphase state exists for $r=2\sqrt{1+\kappa \cos \delta/Z}$.
By substituting expressions for $R(r)$ and $\Theta(r)$ in (\ref{eq:lambda1}) and
(\ref{eq:lambda23}) we find the Floquet exponents for the synchronous solution
\begin{eqnarray}
  \label{eq:lambdavdp1}
  \lambda_2  & = & -\frac{\epsilon}{2} \left( 1+ \frac{\kappa}{Z}\cos \delta \right) \\
  \label{eq:lambdavdp23}
  \lambda_{n,n+1}  & = &  -\frac{\epsilon}{2} \left( 1+ 2\frac{\kappa}{Z}\cos \delta \right)
  \pm \frac{\epsilon}{2} \sqrt{\left( 1+ \frac{\kappa}{Z}\cos \delta \right)^2
        - \left(\frac{\kappa}{Z}\sin \delta \right)^2}.
\end{eqnarray}
In order to test our results we evaluate Floquet multipliers
\footnote{In these comparisons we use Floquet multipliers for
  convenience, since they are easier to handle numerically.  The
  Floquet multipliers are given by $\Lambda_i = e^{2 \pi \lambda_i /
    \omega}+\O(\epsilon^2)$, where $\omega$ is given by (\ref{eq:omega}).}
numerically and compare them to the values estimated from
(\ref{eq:lambdavdp23}). The results are $\O(\epsilon)$ close (Fig.
\ref{fig:vdp}).

\subsection{Van der Pol--Duffing equation}

\begin{figure}
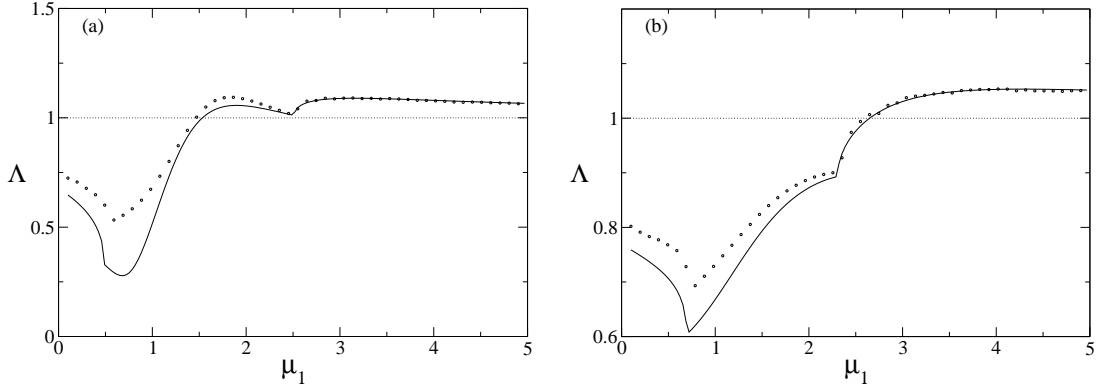

  \centerline{
    \includegraphics[width=7cm]{vdp08D}
    \hspace{0.2cm}
    \includegraphics[width=7cm]{vdp15D}
  }
  \caption{Floquet multipliers for a van der Pol-Duffing array.
    Coupling parameters are $\alpha=-0.2$, $\kappa=1$, $\epsilon=0.1$, and
  (a) $\mu_2=0.8$, (b) $\mu_2=1.5$.}
  \label{fig:vdpD}
\end{figure}

In a recent study of micromechanical and nanomechanical 
resonators models using parametrically driven Duffing oscillators \cite{lifshitz03} and 
van der Pol--Duffing oscillators \cite{cross04} are proposed. 
The van der Pol--Duffing equation
\begin{equation}
\label{eq:dvdp}
        x'' + x - \epsilon (  1 - x^2 ) x' - \epsilon \alpha x^3 = 0,
\end{equation}
is obtained from the van der Pol equation by an addition of a cubic
term. By switching to complex coordinates (\ref{eq:zz}) and writing
(\ref{eq:dvdp}) in normal form we obtain
\begin{eqnarray*}
  z' &=& -iz + \frac{\epsilon}{8}(4z-z^2\bar z + i3\alpha z^2 \bar z) \\
  \bar z' &=& i\bar z - \frac{\epsilon}{8}(-4\bar z +z\bar{z}^2 + i3\alpha z \bar{z}^2).
\end{eqnarray*}
The resonant part of the nonlinearity is given by
$\phi^{\R}(r)=(4-r^2+i3\alpha r^2)/8$. The real part of $\phi^{\R}$ is the same
as in the case of van der Pol oscillator, so this system has the same
inphase solution.  Substitute the real part
$R(r)=1/2-r^2/8$ and imaginary part $\Theta(r) = 3\alpha r^2/8$
of $\phi^{\R}$ in (\ref{eq:lambda1}) and (\ref{eq:lambda23}) to find the
approximate Floquet exponents
\begin{eqnarray}
  \label{eq:lambdavdpD1}
  \lambda_2  & = & -\frac{\epsilon}{2} \left( 1+ \frac{\kappa}{Z}\cos \delta \right) \\
  \label{eq:lambdavdpD23}
  \lambda_{n,n+1} & = &  -\frac{\epsilon}{2} \left( 1 + 2\frac{\kappa}{Z}\cos \delta \right) \\
  & & \pm \frac{\epsilon}{2} \sqrt{\left( 1+ \frac{\kappa}{Z}\cos \delta \right)^2
        - \left(\frac{\kappa}{Z}\sin \delta\right)^2 + 3\alpha \frac{\kappa}{Z}
        (2 \sin \delta + \frac{\kappa}{Z}\sin 2\delta) }.\nonumber
\end{eqnarray}
The numerical simulations (Fig. \ref{fig:vdpD}) support our result.
Note that it follows from (\ref{eq:loadzz}) that
our approximations can be expected to break down for small values of
$\mu_1$.

\subsection{Synchronizing sources and sinks} \label{S:sinks}

\begin{figure}
  \centerline{\includegraphics[width=8cm]{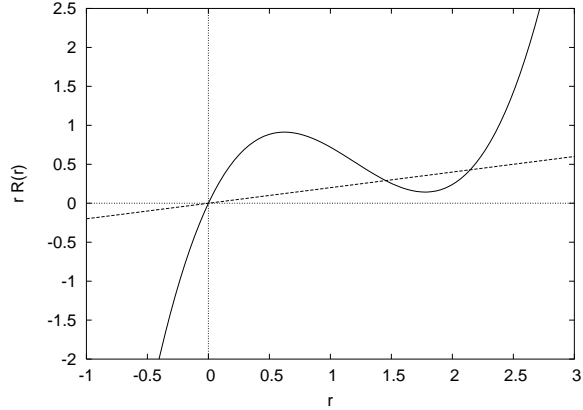}}
  \caption{  
            The radius of the periodic, inphase solution is obtained by 
            solving (\ref{eq:fixed}). For an array of elements (\ref{eq:source})  
            it is given by the intersection of the curve 
            $rR(r) = r/2 (r^2 -3.6 r +3.32)$ and the line 
            $-\kappa \cos \delta/(2Z)r$. If the slope of the line is too 
            small only the trivial solution $r=0$ exists, and the system does not 
            exhibit limit cycle oscillations.}
  \label{fig:inphase}
\end{figure}

\begin{figure}
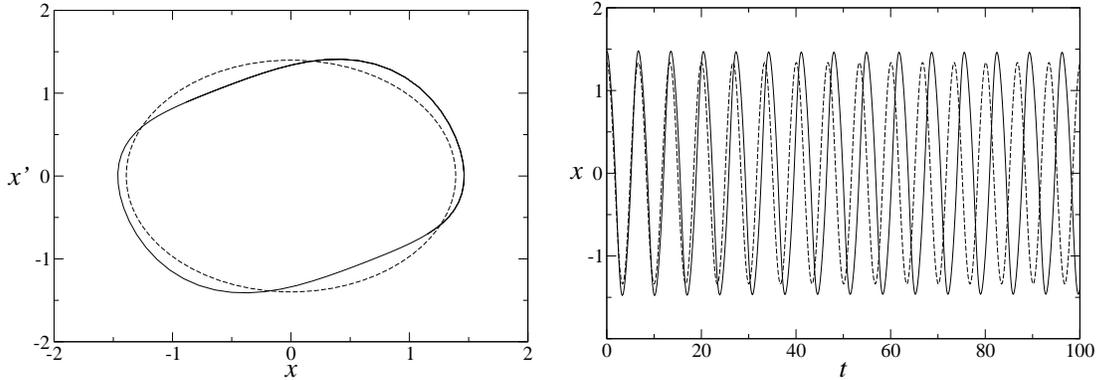

  \centerline{
    \includegraphics[width=7cm]{phaseplot}
  \hspace{0.2cm}
  \includegraphics[width=7cm]{timeplot}}
  \caption{ Phaseplane (left) and time diagram (right) of inphase limit cycle
            solution for coupled sources (\ref{eq:source}). Dashed line represents
            the solution for the truncated normal form system, and solid line
            numerical solution for the full system.}
  \label{fig:phaseplot}
\end{figure}

As noted in Remark \ref{R:sinks}, it is possible to turn a network of
systems with flows that have only a source (or sink) at the origin
into a network of synchronous limit cycle oscillators with an
appropriate coupling.  Consider a dynamical system described by
\begin{equation}
  \label{eq:source}
  x'' +x - \epsilon x' \left[ 4 x^2 \left( 1-\frac{3.6}{\sqrt{x^2+{x'}^2}}
    \right) +3.32 \right] = 0.
\end{equation}
Each individual system has only an unstable fixed point at the origin, and no
other repellers or attractors.  Coupling these systems as in
(\ref{eq:osc}-\ref{eq:load}), and switching to complex variables
(\ref{eq:zz}), gives the following equations of motion
\begin{eqnarray}
  \label{eq:sourcez}
  z' & = & -i z + i \epsilon f(z,\bar z)\\
  \label{eq:sourcezbar}
  \bar{z}' & = & i \bar{z} - i \epsilon f(z,\bar z)
\end{eqnarray}
where the nonlinear term is
\begin{equation}
  f(z,\bar z)= \frac{z-\bar z}{2i}\left[\left( z+\bar z \right)^2
    \left( 1-\frac{3.6}{\sqrt{z \bar z}} \right) +3.32 \right].
\end{equation}
If we keep resonant terms only, (\ref{eq:sourcez}) and (\ref{eq:sourcezbar})
become
\begin{eqnarray}
  z' & = & -i z + \epsilon z \phi^{\R}(z\bar z)\\
  \bar{z}' & = & i \bar{z} + \epsilon \bar z \phi^{\R}(z\bar z)
\end{eqnarray}
with $\phi^{\R} = R(r) = 1/2 (z\bar z -3.6\sqrt{z\bar z} +3.32)$.

\begin{figure}
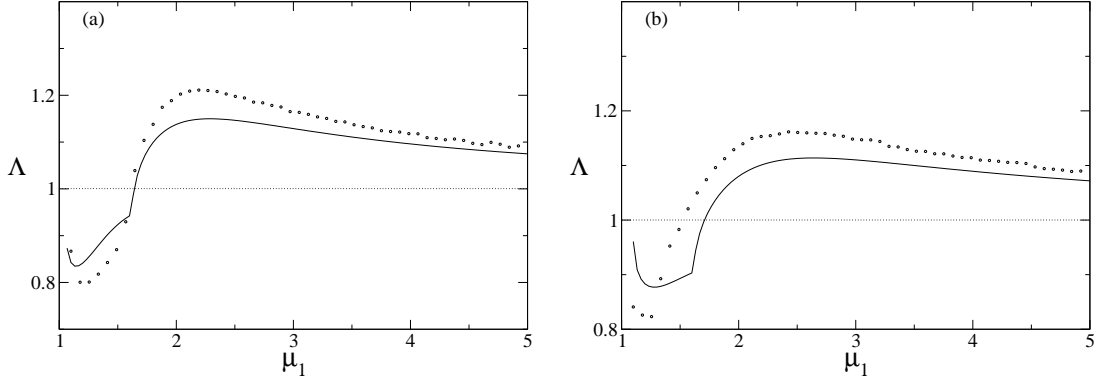

  \centerline{\includegraphics[width=7cm]{vdp08ss-a}
  \hspace{0.2cm}
  \includegraphics[width=7cm]{vdp11ss-a}}
  \caption{Floquet multipliers for coupled sources. $\kappa=1$, $\epsilon=0.1$, and
  (a) $\mu_2=0.8$, (b) $\mu_2=1.1$.}
  \label{fig:ss}
\end{figure}

From (\ref{eq:fixed}) we find that the limit cycle solution does not exist for 
$-\kappa \cos \delta/(2Z) < 0.08$. Below that value the coupled
elements behave as unstable foci, which can be easily checked
numerically. As $-\kappa \cos \delta/(2Z)$ increases above $0.08$ 
the system undergoes a supercritical saddle-node bifurcation of limit cycles, in which 
a stable and an unstable 
inphase limit cycle are created. These two solutions are 
are represented as the intersection of the
curve $R = r/2 (r^2 -3.6 r +3.32)$ and the line $R = -\kappa \cos \delta/(2Z)r$ 
in Fig. \ref{fig:inphase}. 
For a suitable choice of coupling parameters it is possible to obtain 
inphase limit cycle solutions for the array of sources. 
In Fig. \ref{fig:phaseplot} we show oscillations of an 
element in the array, when the coupling
parameters are set to $\kappa=1$, $\mu_1=1.2$ and $\mu_2=0.8$. 
The Floquet exponents for
both limit cycles are readily obtained from (\ref{eq:lambda23}).
These results agree with numerical calculations to the
expected error. In Fig. \ref{fig:ss} we present results for the
``stable'' cycle.

We note that a direct extension of the method discussed in the previous sections was
used to derive these results, since the
nonlinear term in (\ref{eq:source}) is not a polynomial (see \cite{mitropolsky95}).

\section{Conclusion} \label{S:conclusion}

In our study of synchrony in globally coupled oscillator arrays we introduce
a normal form based method, which has 
a number of advantages over methods commonly found in the physics literature. 
The method provides a clear and mathematically rigorous way of finding the 
onset of synchronization in the array with respect to changes in the control 
parameters. It allows for an  easy distinction between 
contributions to the dynamics coming from the network configuration and 
those coming from the internal structure of the network elements. 
As a consequence, we are able to carry out calculations for particular network 
architectures without having to specify the exact form of the nonlinearities 
in the individual elements. 
      
To apply these results to specific weakly nonlinear oscillators it is only necessary
to find the function $\phi^{\R}$, which characterizes the resonant part of the
nonlinearity, and substitute it in the general solution thus obtained. 
The function $\phi^{\R}$ does not depend on the coupling scheme, and is easily 
derived from the equations for the uncoupled
systems. It is therefore tempting to think of synchronization {\em classes} as
different systems may lead to the same $\phi^{\R}$.
  
Although we have chosen a very specific linear coupling in our exposition,
the method can be applied to a variety of 
network configurations simply by retracing the steps we outlined. A similar
calculation can be carried out even if there is a weak nonlinearity in
the coupling equation (\ref{eq:load}) itself. Moreover, the method can be extended to 
higher orders in the small parameter in a 
straightforward fashion, and the procedure can be automated.

The analysis of the synchronous solution of the network
(\ref{eq:osc}-\ref{eq:load}) was particularly simple due its $S_N$
symmetry.  When the network has less symmetry, or only local
symmetries (see \cite{marty}), a similar reduction can be performed.
In such networks one expects polysynchronous solutions, in which groups
of oscillators within the network oscillate synchronously.  One can
further expect to obtain a pair of equations for each cluster of
oscillators in the network \cite{ashwin}.  The stability of these
clusters can then be analyzed in a manner similar to the one
introduced in this paper.

Although we have not treated the case of Josephson junctions, a
similar analysis can also be performed.  In fact we believe that
coherent behavior in networks of various dynamical systems can be
studied using the method of normal forms, and intend to investigate this in the
future.

\ack
SP wants to thank Kurt Wiesenfeld of Georgia Institute of Technology for a number
of fruitful discussions.  The authors thank Arkady Pikovsky for reading 
an earlier version of the manuscript and providing useful comments. 
SP is supported in part by The Office of Naval Research, under grant number N00014-99-1-0592.
KJ was supported in part by grant NSF-0244529.

\section*{References}

\bibliographystyle{unsrt}
\bibliography{refs3}

\end{document}